\documentclass{article}

\usepackage[nonatbib, final]{neurips_2021}

\usepackage[utf8]{inputenc} % allow utf-8 input
\usepackage[T1]{fontenc}    % use 8-bit T1 fonts
\usepackage{hyperref}       % hyperlinks
\usepackage{url}            % simple URL typesetting
\usepackage{booktabs}       % professional-quality tables
\usepackage{amsfonts}       % blackboard math symbols
\usepackage{nicefrac}       % compact symbols for 1/2, etc.
\usepackage{microtype}      % microtypography
\usepackage{xcolor}         % colors

\usepackage{amsmath}
\usepackage{amssymb}
\usepackage{mathtools}

\usepackage[newfloat, frozencache=true,cachedir=_minted-ExternalOperator_NeurIPS2021]{minted}
\usepackage{caption}

\usepackage[ruled,vlined]{algorithm2e}

\usepackage{tcolorbox}
\tcbuselibrary{minted,breakable,xparse,skins}

\definecolor{bg}{gray}{0.95}
\DeclareTCBListing{mintedbox}{O{}mO{}}{%
	breakable=true,
	listing engine=minted,
	listing only,
	minted language=#2,
	minted style=default,
	minted options={%
		linenos,
		gobble=0,
		breaklines=true,
		breakafter=,,
		fontsize=\small,
		escapeinside=||,
		numbersep=8pt,
		tabsize=2,
		encoding=utf8,
		#1},
	boxsep=0pt,
	left skip=0pt,
	right skip=0pt,
	left=25pt,
	right=0pt,
	top=3pt,
	bottom=3pt,
	arc=5pt,
	leftrule=0pt,
	rightrule=0pt,
	bottomrule=2pt,
	toprule=2pt,
	colback=bg,
	colframe=orange!70,
	enhanced,
	overlay={%
		\begin{tcbclipinterior}
			\fill[orange!20!white] (frame.south west) rectangle ([xshift=20pt]frame.north west);
	\end{tcbclipinterior}},
	#3}

\newcommand{\reels}{I\!\!R} %real numbers
\newcommand*\diff{\mathop{}\!\mathrm{d}}

\newtheorem{definition}{Definition}

\newtheorem{example}{Example}

\newcommand{\norm}[1]{\left\lVert#1\right\rVert}

\title{Escaping the abstraction: a foreign function interface for the Unified Form Language [UFL]}

\author{%
  Nacime Bouziani\\%\thanks{email: nacime.bouziani@gmail.com, https://www.imperial.ac.uk/people/n.bouziani18} \\
  Department of Mathematics\\
  Imperial College London\\
  London, UK \\
  \texttt{n.bouziani18@imperial.ac.uk} \\
  \And
   David A.~Ham \\
   Department of Mathematics\\
   Imperial College London\\
   London, UK \\
   \texttt{david.ham@imperial.ac.uk}\\
}

\begin{document}

\maketitle

\begin{abstract}
	High level domain specific languages for the finite element method underpin high productivity programming environments for simulations based on partial differential equations (PDE)  while employing automatic code generation to achieve high performance. However, a limitation of this approach is that it does not support operators that are not directly expressible in the vector calculus. This is critical in  applications where PDEs are not enough to accurately describe the physical problem of interest. The use of deep learning techniques have become increasingly popular in filling this knowledge gap, for example to include features not represented in the differential equations, or closures for unresolved spatiotemporal scales. We introduce an interface within the Firedrake finite element system that enables a seamless interface with deep learning models. This new feature composes with the automatic differentiation capabilities of Firedrake, enabling the automated solution of inverse problems. Our implementation interfaces with PyTorch and can be extended to other machine learning libraries. The resulting framework supports complex models coupling PDEs and deep learning whilst maintaining separation of concerns between application scientists and software experts.
\end{abstract}

\section{Introduction}
\label{sec:Intro}

The design of efficient and composable software relying on high-level languages and automatic differentiation (AD) tools is a major and rapidly growing aspect of modern scientific computing and is of great interest for the machine learning (ML) community. The development of such software requires a vast range of knowledge spanning several disciplines, ranging from applications expertise to mathematical analysis to high performance computing and low-level code optimisation. Software projects relying on automatic code generation have grown in prominence, as their design enables a separation of concerns, increases productivity, and facilitates the collaboration between scientists with different specialisations. Examples of such projects include: Firedrake \cite{Firedrake_2017}, FEniCS \cite{Fenics_2012} and FreeFEM++ \cite{FreeFem++_2012}, in the domain of finite element methods (FEM); and the ML frameworks PyTorch \cite{PyTorch}, TensorFlow \cite{abadi_tensorflow_2016}, Theano \cite{the_theano_development_team_theano_2016} and MXNet \cite{chen_mxnet_2015}. The Unified Form Language (UFL) \cite{UFL_2014} is a domain specific language (DSL) embedded in Python for the finite element method forming the mathematical programming interface of the Firedrake and FEniCS finite element systems. UFL equips Firedrake and FEniCS with a highly expressive interface to specify the variational forms of PDEs and discrete function spaces, providing the abstractions needed for code generation. Finite element problems are solved using gradient-based methods to minimise the problem residual. Consequently, UFL is a fully differentiable language, with inbuilt automatic differentiation.

Domain-specific languages for PDEs are by definition very specific to partial differential equations. The models of real phenomena of interest to scientists and engineers are, regrettably, seldom so straightforward. Alongside the fundamental physical laws expressed as PDEs are various empirical parametrisations, closures, and regularisation terms which represent aspects of the system for which a more fundamental model is either not known or is practically infeasible for some reason.
This scenario is ubiquitous in science and engineering, ranging from geoscience \cite{L_Zanna_2019, lewis_deep_2017, shi_deep_2020, asnaashari_regularized_2013, zhang_regularized_2019} to structural mechanics \cite{rajagopal_implicit_2003, farrell_numerical_2019, oishi_computational_2017} to name but two fields. In order to extend the high productivity, high performance capabilities of UFL and Firedrake to these scenarios, we introduce a foreign function interface to UFL in the form of \textit{external operators}, by which we mean any operator which is not directly expressible in the vector calculus notation of the existing UFL language.

Machine learning (ML) is a natural tool for creating empirical model components from observed data, and provides the motivating example for the external operator concept. ML also demonstrates the criticality of a differentiable programming approach, since backpropagating the neural network involves the differentiation of both the neural net itself and the PDE to which it is coupled. The work presented here allows deep learning frameworks to be coupled directly into Firedrake as external operators, creating common environment for developing PDE models with deep learning components. The external operator feature composes seamlessly with the dolfin-adjoint library \cite{farrell_automated_2013, dolfin-adjoint_2019} enabling automatic differentiation. It interfaces with the PyTorch library and its automatic differentiation engine (\emph{torch.autograd}), and interfaces to other ML frameworks would be straightforward.

\section{The Unified Form Language (UFL)}
\label{sec:UFL}

The Unified Form Language (UFL) \cite{UFL_2014} is an embedded domain-specific language in Python which provides symbolic representations of finite element simulations. UFL forms are compiled by a domain-specific compiler, which takes the high-level description of the weak form of PDEs provided by UFL and translates this representation into low-level code that assembles the sparse matrices and vectors of the finite element problem.

\subsection{Forms}

UFL is organised around representing finite element variational forms, and in particular multilinear forms. A \emph{multi-linear form} (or \emph{linear k-forms}) is a map from the product of a given sequence $\left\lbrace V_{j} \right\rbrace_{j=0}^{k-1}$ of function spaces to a space $K$:

\begin{equation}
\label{def:multilinear_form}
V_{k-1} \times V_{k-2} \times \cdots \times V_{0} \rightarrow K
\end{equation}

that is linear in each argument. The \emph{arity} of a form $k$ is the number of argument spaces. Variational forms with arity $k = 0,\ 1,\ $ and $2$ are respectively named \emph{functionals}, \emph{linear forms} and \emph{bilinear forms}. These can be assembled to produce a scalar, a vector and a matrix, respectively. The UFL variational forms can be parametrised by \textit{coefficient functions}. In this case, the form is expressed as a mapping from a product of a sequence $\left\lbrace W_{j} \right\rbrace_{j=1}^{n}$ of \emph{coefficient spaces} and the \emph{argument spaces}:

\begin{equation}
\label{def:multilinear_form}
W_{1} \times W_{2} \times \cdots \times W_{n} \times V_{k-1} \times V_{k-2} \times \cdots \times V_{0} \rightarrow K.
\end{equation}

We refer to \eqref{def:multilinear_form} as a linear $k$-form with $n$ coefficients. While the multilinear form is necessarily linear in each argument, it can be nonlinear in each coefficient function. An argument is an unknown function in a finite element space, while a coefficient is a known function in a finite element space. 

The multilinear forms represented in UFL constitute the weak form of PDEs and comprise a sum of integrals over subspaces of the problem domain. 

\begin{example}
Let $V$ be a suitable function space and $f$ a given function in $V$. The modified Helmholtz problem is given by: find $u \in V$ such that: 

\begin{equation}
\label{example:var_form}
\int_{\Omega} u\, v + \nabla u \cdot \nabla v\ \diff x = \int_{\Omega} f\, v \diff x \quad \forall v \in V.
\end{equation}

The left-hand side of \eqref{example:var_form} is a 2-form, with $u$ and $v$ as the arguments, and the right-hand side is a 1-form with one coefficient $f$ and one argument $v$.
\end{example}

The symbolic representation of forms in UFL is sufficient to generate the code for matrix and vector assembly, however the execution of that code also requires the simulation data: mesh topology and coordinates, and the value of each coefficient at each finite element node. This is provided by finite element frameworks such as Firedrake and FEniCS, which subclass UFL objects and attach problem data to them, as well as orchestrating the (parallel) assembly and solution operations using the assembly code generated from UFL.

%In UFL, BaseForm primarily describes any object that has arguments, either symbolic or assembled (numeric), e.g. Forms.

\begin{listing}
\captionof{listing}{Firedrake code defining the variational forms in \eqref{example:var_form} using UFL. Note the similarity between lines 9 and 10, and \eqref{example:var_form}.}
\label{code:Assemble-Poisson}
\begin{mintedbox}{python}
from firedrake import *
mesh = |\textcolor{blue}{UnitSquareMesh}|(10, 10)
V = |\textcolor{blue}{FunctionSpace}|(mesh, "Lagrange", 1)
u = |\textcolor{blue}{TrialFunction}|(V)
v = |\textcolor{blue}{TestFunction}|(V)

f = |\textcolor{blue}{Function}|(V)

a = (u * v + |\textcolor{blue}{inner}|(|\textcolor{blue}{grad}|(u), |\textcolor{blue}{grad}|(v))) * dx
L = f * v * dx
\end{mintedbox}
\end{listing}

\subsection{The limited scope of UFL}

As a domain specific language, UFL is specialised to the particular application domain it was built for: the symbolic representation of finite element forms. The vector calculus syntax of UFL can represent the weak form of essentially any partial differential equation. However, realistic applications often augment PDEs with terms not readily expressible in vector calculus. In particular, where the physical basis for part of a model is not well understood it may be advantageous to represent this using a neural network trained on suitable data. These terms could not hitherto be represented in UFL.

\section{External operators: opening a closed language}
\label{sec:Extop}

We present an expressive, flexible and powerful interface for incorporating arbitrary operators in UFL, and providing their implementation to Firedrake. This is achieved by defining symbolic \emph{external operator} objects at the UFL level, and equipping them with the appropriate mathematical properties, such as the ability to be differentiated, so that they can be incorporated directly in UFL expressions, and hence be incorporated directly into the operator assembly and problem solution operations provided by Firedrake.

\begin{definition}
\label{def:External_Operator}
Let $(V_{i})_{1\le i\le l}$, $(W_{i})_{1\le i\le k}$ and $X$ be finite element spaces. An \emph{external operator} $N$ mapping $k$ operands to $X$ is defined as

\begin{equation}
\label{def_external_operator_OPERATOR}
\begin{aligned}
N \vcentcolon  W_1 \times \cdots \times W_k \times V_l \times \cdots \times V_{1} &\longmapsto  X \\
 u_1,\ \ldots,\ u_k,\ v_l,\ \ldots,\ v_1\  \quad &\longrightarrow N(u_1, \ldots, u_k; v_l, \ldots, v_1)
\end{aligned}
\end{equation}

or equivalently,

\begin{equation}
\label{def_external_operator_FORM}
\begin{aligned}
N \vcentcolon  W_1 \times \cdots \times W_k \times  V_l \times \cdots V_1 \times X^{*} &\longmapsto \reels \\
 u_1,\ \ldots,\ u_k,\ v_l,\ \ldots,\ v_1,\ v^{*} \quad &\longrightarrow N(u_1, \ldots, u_k; v_l, \ldots, v_1, v^{*})
\end{aligned}
\end{equation}

\noindent 
where $X^{*}$ is the dual space to $X$. That is, the space of bounded linear functionals over $X$. $N$ can be nonlinear with respect to its operands $u_1, \ldots, u_k$ but it is linear with respect to its arguments $v_l, \ldots, v_1, v^{*}$. The numerical evaluation of $N$ is not specified in the UFL language but left to the specific implementation of the external operator $N$.
\end{definition}

The equivalence of  \eqref{def_external_operator_OPERATOR} and \eqref{def_external_operator_FORM} follows from the reflexivity of $X$. That is,
the canonical injection from $X$ into $X^{**}$ is surjective. As a consequence, we can identify $X$ and $X^{**}$, the space of bounded linear functionals on $X^{*}$. In other words, $X \equiv X^{*} \longmapsto \reels$.

Equation \eqref{def_external_operator_OPERATOR} suggests that external operators behave like operators between finite element spaces whereas definition \eqref{def_external_operator_FORM} suggests that they act as forms in the sense that they are multilinear and scalar-valued. There is no contradiction, both points of view are complementary: they are two different ways to see external operator that are reflected in the implementation. Using \eqref{def_external_operator_OPERATOR}, an external operator can be used in a form anywhere where a coefficient or argument could be used, while using \eqref{def_external_operator_FORM} it is possible to have whole form terms which are simply external operators.

We define a symbolic UFL object representing external operators, the \emph{ExternalOperator}, whose implementation is left to be specified in Firedrake by the user. In other words, the user can build their own external operator by subclassing the external operator class and providing an implementation of the evaluation of the operator and its derivatives. In the case of neural networks, that can be achieved by calling the evaluation and backpropagation of the neural network using the machine learning framework considered by the user (e.g. PyTorch or TensorFlow). Since Firedrake is embedded in Python, the coupling to these frameworks is straightforward. The $m$ hyperparameters of the neural net can be represented in UFL as a coefficient in the very simple finite element space $\reels^m$, which is constant in space over the problem domain.

\section{Differentiation}
\label{sec:Differentiation}

Solving a partial differential equation and training a neural network with a PDE constraint are problems that require the evaluation of the gradient of a form. In UFL, the derivative of a form is based on the G\^ateaux derivative. Because external operators can be used in variational forms, we need to extend UFL automatic differentiation rules to handle external operators. \\

Let $N$ be an external operator of the following form:

\begin{equation}
\label{eq:def_N_star}
\begin{aligned}
N \vcentcolon  V \times V \times  X^{*} &\longmapsto \reels \\
 u,\ m, \ v^{*} \quad &\longrightarrow N(u, m; v^{*})
\end{aligned}
\end{equation}

and consider a PDE, defined by the residual form $F$ linear with respect to $v$, such that:

\begin{equation}
\label{residual_form_F}
F(u, m, N(u,m; v^{*});v) = 0 \quad \forall v\in V.
\end{equation}

In order to solve \eqref{residual_form_F} we need to compute the Jacobian of the residual form, i.e. we need to take the G\^ateaux derivative in the direction $\hat{u} \in V$:

\begin{equation}
\label{eq:jacobian_F_formula}
\frac{\diff F(u, m, N(u,m; v^{*}); \hat{u}, v)}{\diff u} = \frac{\partial F(u, m, N; \hat{u}, v)}{\partial u} + \frac{\partial F(u, m, N; \widehat{N}, v)}{\partial N} \cdot \frac{\partial N(u, m; \hat{u}, v^{*})}{\partial u}
\end{equation}
where $\hat{N}\in X$ is an argument (unknown function). In UFL, $\frac{\partial N(u, m; \hat{u}, v^{*})}{\partial u}$ is a new external operator representing the G\^ateaux derivative of $N$. It is evaluated using the derivative evaluation implementation provided by the user. Listing \ref{code:Compute-Jacobians} shows the UFL code defining and differentiating an external operator and a form containing that operator.
\begin{listing}
\captionof{listing}{Compute $\frac{\partial F}{\partial u}$ and $\frac{\partial N}{\partial u}$ using UFL automatic differentiation}
\label{code:Compute-Jacobians}
\begin{mintedbox}{python}
V = |\textcolor{blue}{FunctionSpace}|(...)
u = |\textcolor{blue}{Coefficient}|(V)
m = |\textcolor{blue}{Coefficient}|(V)
v = |\textcolor{blue}{TestFunction}|(V)
uhat = |\textcolor{blue}{TrialFunction}|(V)

# |$N : V \times V \times X^{*} \longmapsto \mathbb{R}$|
#      |$u,\ m,\ v^{*} \ \ \ \longrightarrow N(u, m; v^{*})$|
N = |\textcolor{blue}{ExternalOperator}|(u, m, function_space=X)

# Define a given form F
F = u * N * v * dx

# Symbolically compute the derivative |$\frac{\partial N(u, m; \hat{u}, v^{*})}{\partial u}$|
dNdu = |\textcolor{blue}{derivative}|(N, u, uhat)

# Symbolically compute the derivative |$\frac{\diff F}{\diff u}$|
dFdu = |\textcolor{blue}{derivative}|(F, u, uhat)
\end{mintedbox}
\end{listing}

Automatic differentiation is also employed to compute the gradient in the task of training a neural network constrained by a PDE. In this case, we would have an external operator of the form $N(u, m; v^{*})$ where $u$ would be the input feeding the model and $m$ the parameters. The training would be performed with respect to a cost functional computed from $u$, the solution to the PDE,  \eqref{residual_form_F} in the cost function. The computation of the gradient of the functional with respect to the neural net parameters $m$ requires the solution of the adjoint PDE. This in turn requires taking the adjoint of $\frac{\partial F}{\partial m}$, which by chain rule necessitates taking the adjoint of $\frac{\partial N}{\partial m}$. This last stage is implemented using backpropagation. In fact, evaluating the adjoint of the gradient of the external operator with respect to the parameters will call the evaluation implementation from the external operator subclass representing the model, which itself delegates evaluation to  the machine learning framework in which the model is implemented. At the UFL level, this can be simply achieved via the code in Listing \ref{code:backpropagation-Firedrake}.
\begin{listing}
\captionof{listing}{Assembling $\frac{\partial N}{\partial m}^{*} \cdot y$ in Firedrake}
\label{code:backpropagation-Firedrake}
\begin{mintedbox}{python}
\# Symbolically compute the derivative |$\frac{\partial N}{\partial m}$|
dNdm = |\textcolor{blue}{derivative}|(N, m)

# Symbolic operation: |$\frac{\partial N}{\partial m}^{*}\cdot y$|
backprop_y = |\textcolor{blue}{action}|(|\textcolor{blue}{adjoint}|(dNdm), y)

# Evaluation
|\textcolor{blue}{assemble}|(backprop_y)

\end{mintedbox}
\end{listing}

\section{Example: seismic inversion}

Seismic inversion is the use of seismic reflection data to infer the material properties of the Earth's subsurface. In this process, waves are used to interrogate a medium, and the medium's response is recorded and processed to obtain a description of the subsurface. The problem arises predominantly in exploration geophysics (e.g. oil and gas prospection) \cite{Multiscale_FWI_2013, Seismic_Imaging_adjoing_tomography_2012, Seismic_Tomography_2010} and geotechnical site characterisation \cite{FWI_site_characterization_2013}. This problem can be formulated as an inverse wave problem. This problem is invariably ill-posed: various solution artifacts result in indistinguishable functionals but nonphysical solutions. A simple model problem can be formulated as follows:

\begin{equation}
\label{regularizer_example}
\min_{c \in P}\ \ \frac{1}{2}\norm{\varphi(c) - \varphi^{obs}}_{V}^{2} + \alpha \mathcal{R}(c)
\end{equation}

where $\varphi^{obs}$ are the observed data, $c$ the scalar wave speed, $\alpha$ the regularisation factor and $\varphi \in V$ is the wave displacement of the medium such that:

\begin{equation}
\label{regularizer_wave_Example}
\begin{aligned}
 - \frac{\partial^{2} \varphi}{\partial t^{2}}  - c^{2} \Delta \varphi &= f &\textrm{in}\ \Omega\\
 	\varphi(t=0) = \varphi_0.
\end{aligned}
\end{equation}

Equation \eqref{regularizer_wave_Example} is referred to as the \emph{forward problem}. In practice, one may use a more complex formulation of this equation to take into account additional physical effects such as elasiticy, acousticity or anisotropy. More realistic applications may also have complex boundary conditions.

Equation \eqref{regularizer_example} is the function to optimise. The first term accounts for the mismatch error between the solution obtained by the forward problem (the PDE) and the observed data. We refer to this term as the \emph{fidelity term}. $\mathcal{R}$ is a regularisation operator which attempts to counter from noisy observations and to help making the problem well-posed. The PDE is based on fundamental physical laws but the regulariser is more heuristic in nature. A number of recent works have employed deep learning to build regularisers for inverse problems, especially for seismic inversion \cite{lewis_deep_2017, shi_deep_2020, zhang_regularized_2019} and medical imaging \cite{lunz_adversarial_2018, li_nett_2019, adler_solving_2017}. External operators in UFL provide a high-level programming abstraction for this problem. The cost function simply contains an external operator representing the neural network model, i.e. $\mathcal{R}(c) = \frac{1}{2}\norm{N(c)}^{2}$, with $N$ an external operator. Listing \ref{code:seismic-inversion} illustrates this in Firedrake.

\begin{listing}[h!]
\captionof{listing}{Outline of seismic inversion using a neural net regulariser implemented as an ExternalOperator in Firedrake.}
\label{code:seismic-inversion}
\begin{mintedbox}{python}
from firedrake import *
from firedrake_adjoint import *

...
# Get a pre-trained PyTorch model
model = ...
# Define the external operator from the model
pytorch_op = |\textcolor{blue}{neuralnet}|(model, function_space=...)
N = pytorch_op(vel)

# Solve the forward problem defined by equation |\eqref{regularizer_wave_Example}|
phi = F(c)
# Assemble the cost function: |$J = \frac{1}{2} \norm{\varphi(c) - \varphi^{obs}}^{2} + \frac{\alpha}{2} \norm{N(c)}^{2} $|
J = |\textcolor{blue}{assemble}|(0.5*(|\textcolor{blue}{inner}|(phi-phi_obs, phi-phi_obs) +
     					   alpha*|\textcolor{blue}{inner}|(N, N))*dx)

# Optimise the problem
Jhat = |\textcolor{blue}{ReducedFunctional}|(J, |\textcolor{blue}{Control}|(c))
c_opt = |\textcolor{blue}{minimize}|(Jhat, method="L-BFGS-B", tol=1.0e-7,
							   options={"disp": True, "maxiter" : 20})
\end{mintedbox}
\end{listing}

\begin{figure}[htp]
\centering
\includegraphics[width=.30\linewidth]{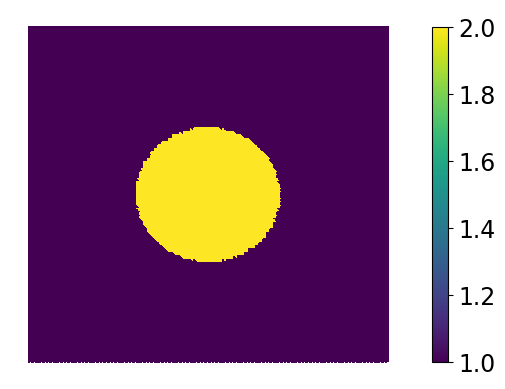}
\includegraphics[width=.32\linewidth]{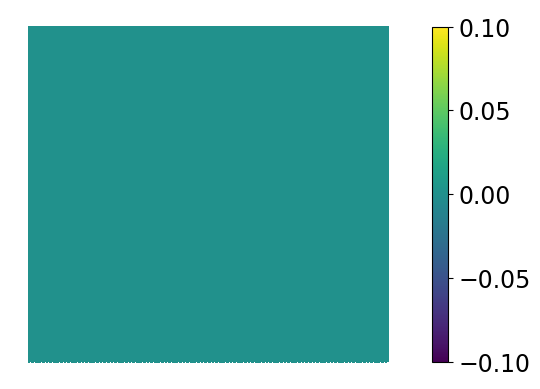} \\
\includegraphics[width=.32\linewidth]{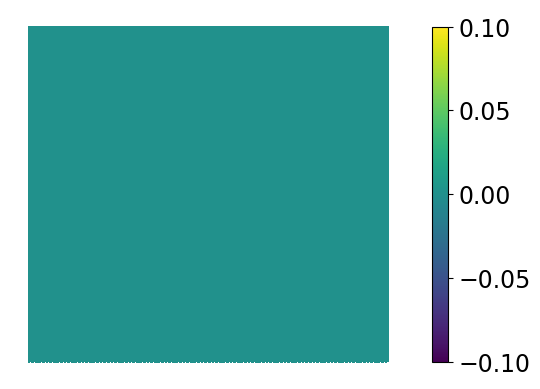}
\includegraphics[width=.30\linewidth]{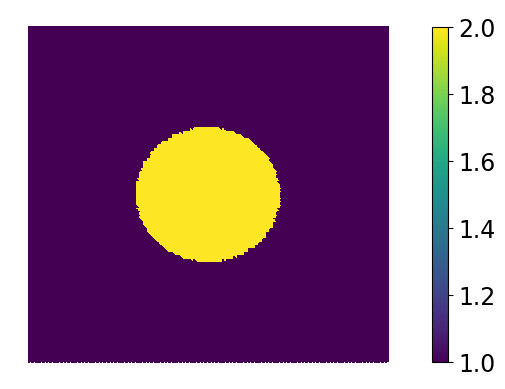}
\caption{Recovered wave speed $c$ as a function of position ($x,z$) obtained for the model \eqref{regularizer_example}-\eqref{regularizer_wave_Example}: exact velocity (upper left), without regularisation (upper right), Tikhonov regulariser (lower left), neural network-based regulariser (lower right).}
\label{fig:velocity_models_seismic_inversion}

\end{figure}

The exact version of Firedrake (including UFL) used in this paper is archived in \cite{zenodo/Firedrake-20210924.0} while the code used to run the example and generate the output figure is archived in \cite{Bouziani_Seismic_inversion_using_2021}

\subsection*{Acknowledgement}

This work was funded by a President’s PhD Scholarship at Imperial College London.

\bibliographystyle{plain}
\bibliography{bibliography.bib}

\end{document}